\documentclass[twocolumn,showpacs,showkeys,pra]{revtex4-2}
\usepackage{amsmath}
\usepackage{graphicx}
\usepackage{color}
\usepackage{amssymb}
\usepackage{float}

\begin{document}
\title{	
Josephson effect and self-trapping in helicoidal spin-orbit coupled Bose-Einstein condensates with optical lattices}
\author{Sumaita Sultana}
\email{reach2sumaita@gmail.com}
\affiliation{Department of Physics, Kazi Nazrul University, Asansol-713340, W.B., India}
\author{Golam Ali Sekh}
\email{skgolamali@gmail.com( corresponding author)}
\affiliation{Department of Physics, Kazi Nazrul University, Asansol-713340, W.B., India}
\begin{abstract}
We consider matter-wave bright solitons in helicoidal spin-orbit coupled Bose-Einstein condensates in optical lattices with a view to study Josephson-type oscillation and self-trapping of population imbalance between two pseudo-spin states. For fixed values of nonlinear interaction and lattice parameters, the population imbalance executes Josephson-type oscillation and the frequency of oscillation decreases with the increase of helicoidal gauge potential. The frequency of oscillation, however, increases in presence of nonlinear optical lattices. We find that the population imbalance  can oscillate about a non-zero value and result self-trapping for proper choices of parameters of the system. The self-trapping can be enhanced by the increase of helicoidal gauge potential and by properly tuning the lattice parameters.
\end{abstract}

\keywords{Helicoidal spin-orbit coupling; Bose-Einstein condensates; Josephson oscillation; Linear and nonlinear optical lattices; Quantum mechanical self-trapping}
\pacs{05.45.Yv,03.75.Lm,03.75.Mn}
\maketitle
\section{Introduction}
Josephson physics describes the tunneling phenomena that arise due to  coupling between two macroscopically phase coherent quantum states such as superconductors separated by thin insulating layer. {Due to this tunneling effect a constant supercurrent flows through the insulating layer in the case of d.c. Josephson effect while an oscillating supercurrent flows if a constant voltage is applied in the case of a.c. Josephson effect. Here the constant voltage induces temporal phase in suppercurrent. In weakly coupled Bose-Einstein condensates~(BEC) separated by tunneling layer if a constant chemical potential difference is applied then the interplay between tunneling and interaction causes the oscillating atomic-supercurrent to flow  and thus results an oscillation of atomic population difference between these two coupled BECs \cite{r1,r2}. This is termed as Josephson-type oscillation.} If interaction dominates over the tunneling, a transition from Josephson oscillation to self-trapping occurs resulting the population imbalance to oscillate about a non-zero value. This tunneling oscillation can be explored in building quantum mechanical circuit to design qubit \cite{r3}, SQUIDS \cite{r4} and to engineer quantum states \cite{r5}.

A spin-orbit coupled (SOC) atomic BEC created by artificial gauge potentials offers a new possibility to study Josephson oscillation in pseudo-spin 1/2 systems. This type of oscillation in a weakly interacting BEC is a nonlinear generalization of the typical a.c. and d.c. Josephson oscillations. Thus it has received special attention and resulted several interesting experimental \cite{r6, r7, r8} and theoretical \cite{r9, r10, r11, r12, r13, r14, r15, r16, r17} investigations. Particularly, theoretical investigations in this direction include the studies on Josephson dynamics in two species BECs in double-well potential \cite{r10}, internal Josephson oscillation between coupled solitons with time varying Raman frequency \cite{r17}, effects of SOC on Josephson tunnelling between two annular condensates \cite{r18}, Josephson oscillation of 2D chiral solitons \cite{r19} and Josephson-type oscillation in presence of optical lattices \cite{r20}.

Besides the above mentioned Josephson effects in real space, it has also been predicted theoretically \cite{r21} and, realised experimentally \cite{r22} momentum-space Josephson effect in SOC-BEC. Studies in this direction are not only limited to atomic-bosonic systems but also extended from exciton-polariton systems \cite{r23} to fermionic superfluid gases \cite{r24, r25}.

Various techniques for generating artificial gauge potentials allow to tune strength and shape of spin-orbit coupling in BECs \cite{r26, r27, r28}. {For the appropriate choice of gauge potential, ${\bf A}(x)=\boldsymbol{\sigma}\cdot \mathbf {n}(x)$ with $\mathbf{n}(x)=(\cos(2\beta x),\sin(2\beta x),0)$ and Pauli matrices ($\boldsymbol{\sigma}$)$=(\sigma_1,\sigma_2, \sigma_3)$,} one can induce $\pi/\beta$ periodic helicoidal non-uniform spin-orbit coupling  in BECs.
{The periodic SO coupling has been considered to find its influence on the band structure, wave functions, and spin-density distribution of two-dimensional electron gas \cite{2de}. The so-called photonic graphene has been created through the generation of evanescently coupled helical waveguides where the propagation of light is described by vector potential \cite{hwg}}.

{The helicoidal shaped vector potential ${\bf A}(x)$ introduces point translational symmetry in the system. However, one can  switch  to a  rotating frame using an appropriate gauge transformation for a chosen gauge potential ${\bf A}(x)$ , where the point translational symmetry of the system transform into continuous translational symmetry}.  {In this frame,  the coefficients of the resulting  equations become space-independent and  the SOC becomes non-uniform.}
This non-uniform SOC is highly tunable and can increase the effects of intrinsic nonlinearity \cite{r26, r27, r28, r29, r30, r31}. A Bose-Einstein condensate with helicoidal SOC features many properties of integrable system such as stable propagation of bright solitons \cite{r32}, Peregrine solitons \cite{r33}, bright-dark Peregrine solitons \cite{r34}, stripe solitons \cite{r35}, multi-hump solitons \cite{r36} and kink solitons \cite{r37}. One of the important aspects of the soliton in helicoidal SOC-BECs is that the soliton with any peak amplitude is non-radiative and thus can propagate a large distance \cite{r32}. In a recent study it is shown that the helicoidal gauge potential can be used to manipulate modulational instability in BECs \cite{r38}.

Our objective in this paper is to study Josephson-type~(JT) oscillation and self-trapping of atomic population between two solitons in helicoidal spin-orbit coupled quasi-one-dimensional~(Q1D) BECs with optical lattices using an effective one-dimensional(1D) Gross-Pitaevskii equation(GPE). Particularly, we see that the atomic imbalance between the two solitons oscillates periodically and the frequency of oscillation depends sensitively on the strength of helicoidal gauge potential. The frequency decreases with the increase of helicoidal gauge potential. In presence of optical lattices, the frequency of oscillation further increases. We see that there exists a critical point for the observation of self-trapping. Interestingly, the critical point gets shifted in the parameter space with the change of helicoidal gauge potential and Zeeman-splitting. { We note that  there is no the so-called Rabi coupling, a linear coupling between the wave functions of the condensates,  in helicoidal SOC-BECs. Here the two states are coupled through the gradient of the wave function i.e., helicoidal gauge potential. Therefore, we say that the oscillation of population imbalance can be affected by the helicoidal SOC.} {Recently, we consider uniform spin-orbit coupled BECs  \cite{r39, r40, r41} in presence of nonlinear optical lattices (NOLs) and see that our approach can catch the effects of the NOLs and Rabi coupling on JT oscillation and self-trapping \cite{r20}. Here we extend the same to the case of non-uniform SOC-BECs with a view to extract the effect of helicoidal gauge potential on the system.}

In section II, we present a theoretical model to describe helicoidal SOC-BEC in optical lattices. Based on a variational approach, we derive equations for the center of mass, population imbalance and phase difference of the coupled condensates. In section III, { we first fix the center of mass of soliton pair in a particular position in the linear optical lattice(LOL). Here the LOL acts as a trap. We then study the effects of periodic modulation of atomic interaction or the so-called nonlinear optical lattice (NOL)  on the JT oscillations for different strengths of the helicoidal gauge potentials.}  In section IV, we find a condition for the observation of quantum mechanical self-trapping and its dependence on the helicoidal gauge potential and nonlinear optical lattice. Finally, we devote section V to summarize the results of our investigations.
\section{General Formalism}
A Bose-Einstein condensate can be treated theoretically by a single mean-field Gross-Pitaevskii equation. In presence of helicoidal spin-orbit coupling, the BEC describes a pseudo-spin 1/2 system and thus it becomes a binary BEC. This system in quasi-one-dimensional (Q1D) geometry is modelled by the following coupled Gross-Pitaevskii equations (GPE) \cite{r32, r38}.
\begin{eqnarray}
i\frac{\partial \psi_j}{\partial t}
&=&-\frac{1}{2}\frac{\partial^2\psi_j}{\partial x^2}-i\beta(-1)^j \frac{\partial\psi_j}{\partial x}-i\alpha_s\frac{\partial\psi_{3-j}}{\partial x}+V_{ext}(x)\psi_j \nonumber\\ &-& \frac{\Delta}{2}(-1)^j\psi_j-(g_1|\psi_j|^2+g_{2}|\psi_{3-j}|^2)\psi_j,
\label{eq1}
\end{eqnarray}
with the external potential
\begin{eqnarray}
V_{ext}(x)=\frac{1}{2}{({\omega_x}/{\omega_\perp})^2 }x^2+V_0 \cos(2k_l x).
\label{eq3}
\end{eqnarray}
Here, $\psi_j(x,t)$ $(j=1,2)$ gives the order parameter. {The strengths of attractive intra- and inter-atomic interactions are represented by $g_1$ and $g_2$. Here we take same intra-atomic interaction for both the pseudo-spin states.}  In Eq. (1), the parameters $\alpha_s $, $\beta$ and $ \Delta $ stand for the strengths of spin-orbit coupling, helicoidal gauge potential (HGP) and Zeeman splitting respectively. The external potential in Eq. (\ref{eq3}) contains two terms, namely, harmonic trapping potential (first term) and linear optical lattice (second term) with wavenumber $k_l$ and strength $V_0$. We make Eq. (1)  dimensionless by re-scaling the energy, length and time of the system in the units of  $\hbar \omega_\perp $, $a_\perp=\sqrt{{\hbar}/{m\omega_\perp}}$ and $({\omega_\perp})^{-1}$ respectively. Here $\omega_x$ and $\omega_\perp$ stand for the angular frequencies of the harmonic trap in the axial and transverse directions.

With a view to consider a helicoidal spin-orbit coupled BEC trapped in linear optical lattices (LOL), we take $\omega_\perp \gg \omega_x$ such that  the effects of harmonic trap is negligible. In addition, we introduce periodicity in the system by modulating the nonlinear mean-field interaction periodically, called nonlinear optical lattice (NOL). The NOLs are given by \cite{nol1,nol2,nol3}
\begin{subequations}
\begin{eqnarray}
g_1=\gamma_0+\gamma_1 \cos(2k_n x),
\label{eq4a}
\end{eqnarray}
\vskip -0.75cm
\begin{eqnarray}
g_{2}=\beta_0+\beta_1 \cos(2k_n x),
\label{eq4b}
\end{eqnarray}
\end{subequations}
where $k_n$ is the wavenumber of the NOLs with their strengths $\gamma_1$ and $\beta_1$ respectively. {This lattice can be engineered  by spatial periodic variation of  either laser intensity in the vicinity of optical Feshbach resonance \cite{bm1,bm2} or magnetic field intensity in the vicinity of optically induced magnetic Feshbach resonance \cite{fr1,fr2}}.   {Generally, the amplitude of periodic modulation is smaller than that of the unmodulated part of interaction such that $\gamma_1 < \gamma_0$ and $\beta_1 < \beta_0$.  Understandably, $\gamma_0>\beta_0$ for miscible binary condensates.} The NOLs are found very effective in controlling the dynamics and stability of matter-wave solitons in BECs \cite{r42, r43,  r45, r46}.
In order to proceed with the variational approach, we recast the Eqs. (1) and (2) in terms of the following Lagrangian density.
\begin{eqnarray}
\mathcal{L}&=&\sum_{j=1}^2\bigg[-\frac{1}{2}\bigg|\frac{d\psi_j}{dx}\bigg|^2+(-1)^j\frac{\Delta}{2}|\psi_j|^2+\frac{\gamma_0}{2}|\psi_j|^4
\nonumber\\&+&
\bigg(\frac{i}{2}\bigg(\psi_j^*\frac {d\psi_j}{dt}+(-1)^j\beta\psi_j^*\frac{d\psi_j}{dx}+{\alpha_s}\psi_j^*\frac{d\psi_{3-j}}{dx}\bigg)+c.c\bigg)
\nonumber\\&+&\frac{\gamma_1}{2}\cos(2k_n x)|\psi_j|^4+\frac{\beta_1}{2} \cos(2 k_n x)|\psi_j|^2|\psi_{(3-j)}|^2
\nonumber\\&+&\frac{\beta_0}{2}|\psi_j|^2|\psi_{(3-j)}|^2 
-V_0 \cos(2k_l x)|\psi_j|^2\bigg].
\label{eq5}
\end{eqnarray}
A helicoidal spin-orbit coupled BEC supports freely moving bright solitons both in presence and absence of Zeeman splitting \cite{r32}. Thus, we adopt the following Gaussian-shaped trial solutions for bright solitons \cite{r17}.
\begin{equation}
\psi_j = \frac{e^{-(x-x_0)^2/2w^2}}{{(\sqrt{\pi} w})^{1/2}}
\sqrt{N_j} e^{ik_j (x-x_0)+i\phi_j}.
\label{eq6}
\end{equation}
Here, $x_0 $, $w $, $k_j $ and $ \phi_j $ ($j=1, 2$) are time-dependent variational parameters which represent respectively, center of mass, width, wavenumber and phase of the soliton solutions. Understandably, {$N=\int\left(|\psi_1|^2+|\psi_2|^2\right)\,dx=(N_1+N_2)$ represents the norm of the system and it gives scaled value of total number of atoms \cite{gercia}.  In this trial solution, we allow  amplitude of the individual component to vary with number of atoms keeping the width constant. The two-component solitons containing atoms of different spins prefer spin-mixed phase due to SU(2) atomic interaction and  thus we take  identical center of mass of both the solitons.}

Substituting Eq. (\ref{eq6}) in Eq. (\ref{eq5}) and then integrating with respect to $x$ from $-\infty$ to $+\infty$, {we get the following   Lagrangian.}
\begin{eqnarray}
L&=&\!\sum_{j=1}^2\bigg[ \!\!-N_j\bigg(\frac{d\phi_j}{dt}-k_j\frac{dx_0}{dt}+\frac{1}{4w^2}+\frac {k_j^2}{2}+(-1)^j\beta k_j\bigg) \nonumber\\&+&\frac{1}{2\sqrt{2\pi }w}(\gamma_0 N_j^2+\beta_0 N_j N_{3-j})-\frac{V_0 N_j}{e^{k_l^2 w^2}}\, \cos(2 k_l x_0)
\nonumber\\&+&\frac{e^{\frac{-k_n^2 w^2}{2}}}{2\sqrt{2\pi}w} (\gamma_1 N_j^2+\beta_1 N_j N_{3-j})\cos(2 k_n x_0)\nonumber\\&+&(-1)^j\frac{\Delta N_j}{2}-\alpha_s\sqrt{N_1N_2}k_je^{-k_-^2w^2}\cos(\varphi)\bigg].
\label{eq7}
\end{eqnarray}
Here $ \varphi=\phi_1-\phi_2 $. Making use of the Ritz  optimization conditions: $\frac{\delta L}{\delta w}=0$, $\frac{\delta L}{\delta N_1}=0$, $\frac{\delta L}{\delta N_2}=0$, $\frac{\delta L}{\delta \phi_1}=0$, $\frac{\delta L}{\delta \phi_2}=0$, $\frac{\delta L}{\delta k_1}=0$ and $\frac{\delta L}{\delta k_2}=0$,
we obtain the following coupled equations.
\begin{eqnarray}
\frac{dx_0}{dt}&=&k_+ +\frac{\alpha_s e^{-\beta^2w^2}}{\sqrt{1-Z^2}}\cos(\varphi),
\label{eq8}\\
\frac{dk_+}{dt}&=&2\beta\alpha_s k_+\sqrt{1-Z^2}e^{-\beta^2 w^2} \sin(\varphi)\nonumber\\&-&\frac{N k_n \sin(2k_n x_0)}{2\sqrt{2\pi}w} e^{\frac{-k_n^2 w^2}{2}}[G_++G_- Z^2]
\nonumber\\&+&2V_0k_le^{-k_l^2w^2}\sin(2k_lx_0),
\label{eq9}\\
\frac{dZ}{dt}&=&-2e^{-\beta^2 w^2}\alpha_s k_+\sqrt{1-Z^2}\sin(\varphi),
\label{eq10}\\
\frac{d\varphi}{dt}&=& 2\beta\dot x_0+\Lambda Z+\frac{NZ}{\sqrt{2\pi}w}e^{\frac{-k_n^2\omega^2}{2}}\cos(2k_n x_0)G_-\nonumber\\&+&\frac{2\alpha_s k_+ Z e^{-\beta^2w^2}}{\sqrt{1-Z^2}}\cos(\varphi)-\Delta,
\label{eq11}
\end{eqnarray}
with $ \Lambda=N(\gamma_0-\beta_0)/(\sqrt{2\pi})w$, $k_\pm=(k_1\pm k_2)/2$, $G_-=\gamma_1-\beta_1$ and $G_+=\gamma_1+\beta_1$. Here, $ \varphi$ and $ Z=(N_1-N_2)/N$ represent respectively phase difference and population imbalance between the two solitons.
{We know that $k_{-}$ is small since $k_1\approx k_2$ due to the assumption of same center of mass of both the solitons.  Therefore, for weak helicoidal potential, we can take $k_-\approx \beta$.}
The coupled Eqs. (7)-(10) show that the rate of change of population imbalance depends directly on the  initial phase difference and SOC parameters. Clearly, its dependence on the lattice and other parameters comes through $\varphi$. Therefore, it will be quite interesting to check how the population imbalance are affected by the different parameters of the system.

\section{Helicoidal gauge potentials and JT oscillation}
Let us consider that the coupled solitons are located at a particular position $(x_0={\rm const.})$ in the linear optical lattice~(LOL). For these stationary solitons, the center of mass of the coupled system does not move in time and thus $\dot{x_0}=0$. In this case, the coupled Eqs. (7)-(10) can be reduced as follows.
\begin{eqnarray}
\frac{dZ}{dt}&=& - A{\sqrt{1-Z^2}} \sin(\varphi),
\label{eq15}\\
\frac{d\varphi}{dt}&=& \Lambda Z+\frac{AZ}{\sqrt{1-Z^2}}\cos(\varphi)+BZ-\Delta.
\label{eq16}
\end{eqnarray}
Here $A=2k_0\alpha_s e^{-\beta^2w^2}$ and $B=\frac{NG_-e^{\frac{-k_n^2w^2}{2}}}{\sqrt{2\pi}w}\cos(2k_n x_0)$.
Note that Eqs. (11) and (12) are obtained by considering weak spin-orbit coupling such that the first term in Eq. (8) is negligible and by choosing the parameters of both the  linear and nonlinear lattices such that 
\begin{eqnarray}
G_-\geq\frac{4}{Z_0^2}\bigg[\frac{\sqrt{2\pi}wV_0}{e^{\frac{w^2}{2}(2k_l^2+k_n^2)}}\frac{k_l\sin(2k_lx_0)}{\sin(2k_nx_0)}-\frac{G_+}{4}\bigg].  
\label{eq17}
\end{eqnarray}
{Here $Z_0$ stands for the initial population imbalance and it lies in the range $0<Z_0<1$.} 
Under this condition $ k_+ $ remains approximately constant with time and {thus we take $k_{+}=k_0$ in Eqs. (\ref{eq15}) and (\ref{eq16})}. 
Therefore, Eqs. (11) and (12) can describe the variation of population imbalance with a change of relative phase between the two spin states in presence of both linear and nonlinear optical lattices (NOL).

With a view to study population imbalance dynamics, we consider that the components are miscible~($\gamma_0>\beta_0$) and solve Eqs. (11) and (12) numerically for different values of initial phase differences and display the results in Fig. 1. We see that, in absence of NOLs the population imbalance oscillates with time and the amplitude of oscillation depends sensitively on the initial value of $\varphi$ (left panel). More specifically, the amplitude increases with the increase of $\varphi(0)$ for a fixed value of $\beta$. This type of oscillation is often termed as Josephson-type~(JT) oscillation \cite{r16, r20, r17}. The  frequency of oscillation decreases with the increase of $\beta$~(right panel)  and thus we say that population imbalance dynamics can be decelerated  with the help of helicoidal gauge potential.
\begin{figure}[h!]
\includegraphics[scale=.30]{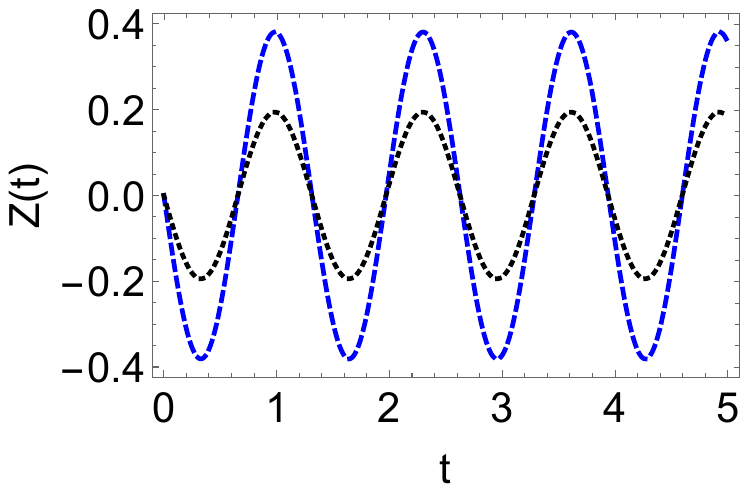}
\includegraphics[scale=.30]{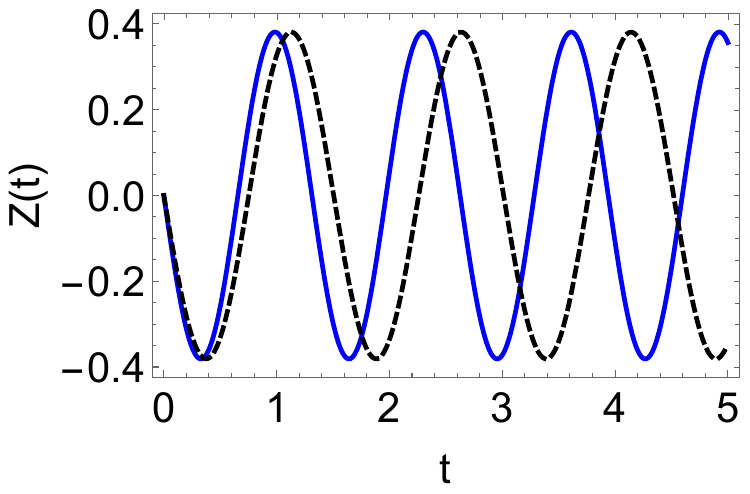}
\caption{Left panel: Variation of population imbalance with $t$ in absence of nonlinear optical lattices for $\beta=1.6$. Here, dashed blue and dotted black are drawn for the initial phase differences $\varphi(0)= \pi/8$ and $\pi/16$ respectively. Right panel: It gives $Z(t)$ versus $t$ for $\beta=1.6$ (solid blue) and $\beta=2$ (dashed black) with $\varphi(0)= \pi/8$. In both the panels, we take $\alpha_s=4$, $x_0=0.25$, $\Delta=0$,  $B=0$, $k_0=1$ and $\Lambda= 0.04$.}
\label{fig1}
\end{figure}
\begin{figure}[h!]
\includegraphics[scale=.30]{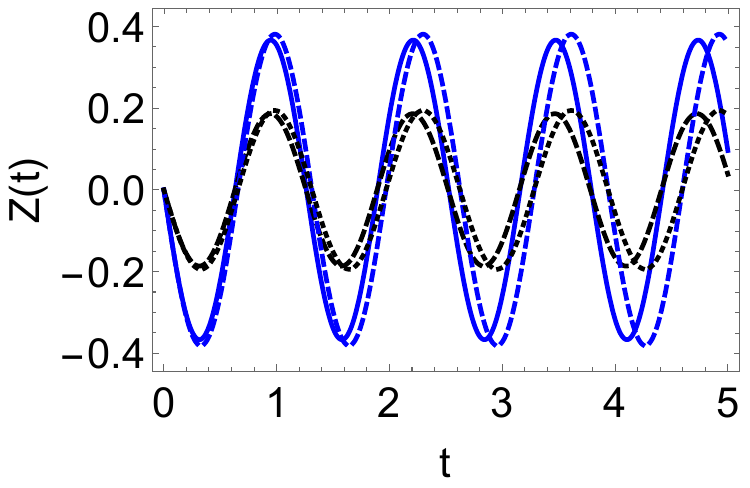}
\includegraphics[scale=.30]{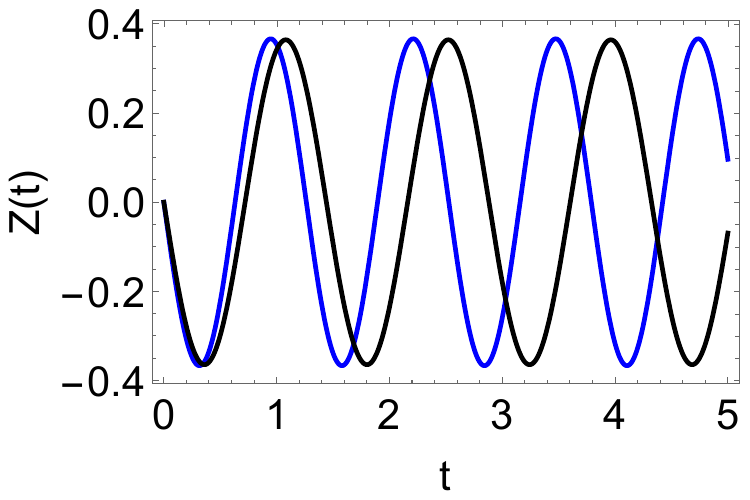}
\caption{{Left panel: Variation of population imbalance with t in presence of NOL for $\beta=1.6$. Here, solid-blue and dashed- black curves are drawn for the initial phase differences  $\varphi(0)= \pi/8$ and $\varphi(0)= \pi/16$ respectively in presence of NOL while the dashed-blue and dotted-black curves are in absence of nonlinear optical lattices.} Right panel: It gives $Z(t)$ versus $t$ for $\beta=1.6$ (solid-blue) and $\beta=2$ (solid-black) with $\varphi= \pi/8$. In both the panels, we take $\alpha_s=4$, $x_0=0.25$, $G_-=1.05$, $V_0=2$, $k_l=5$, $k_n=1.8$, $\Delta=0$, $k_0=1$ and $\Lambda=0.04$.}
\label{fig2}
\end{figure}

\begin{figure}[h!]
\includegraphics[scale=.30]{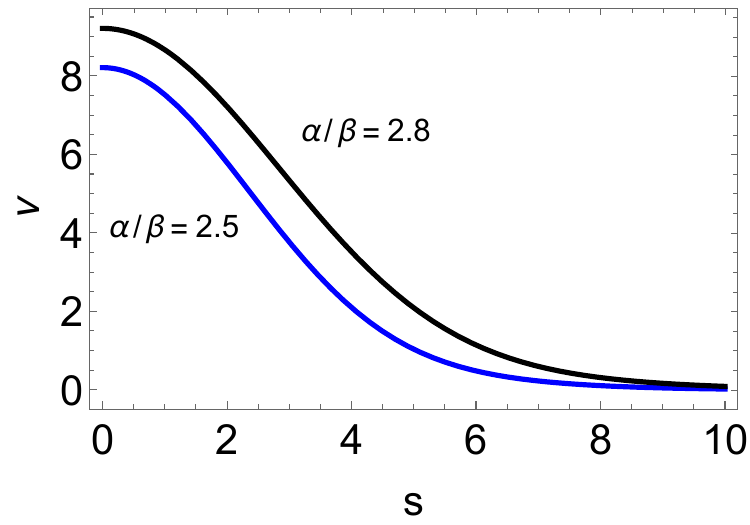}
\includegraphics[scale=.32]{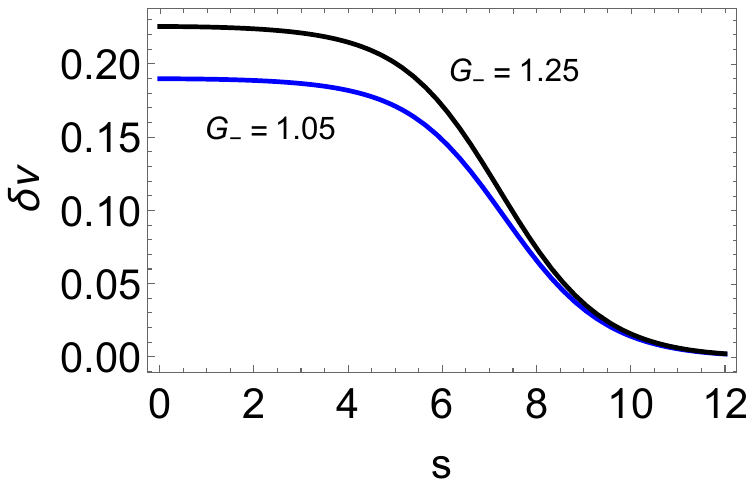}
\caption{Left panel: Variation of JT oscillation frequency with $s$. Here $G_-=1.05$. Right panel: Variation of  difference ($\delta\nu$) between the frequencies in presence and absence of  the NOL  with $s$. In both the panels, we take $x_0=0.25$, $\Lambda=0.04$, $\alpha_s=4$, $k_l=5$ and $k_n=1.8$.}
\label{fig3}
\end{figure}

{The effects of NOL on the JT oscillation in helicoidal SOC-BEC is shown in Fig. 2 taking $k_n< k_l$. Here the LOL  is incommensurable with the NOL.} We see that the $Z(t)$ oscillates with time for different $\varphi(0)$ and the oscillations are quite similar to those observed in absence of NOLs. However, the frequency of oscillation~($\nu$) is larger than that found in absence of NOL. This implies that the NOL causes acceleration in the dynamics of population imbalance.  {We have checked that the numerical value of $\nu$ decreases as $k_n\rightarrow k_l$.}

The variation of $\nu$ for different lattice and SOC parameters for small oscillations can approximately be calculated from the relation given below.
\begin{eqnarray}
\nu = [A^2+A(\Lambda+B)]^\frac{1}{2}.
\label{eq15a}
\end{eqnarray}
In writing Eq. (\ref{eq15a}) we have used $\ddot{Z}+\nu^2 Z=0$. Understandably, $\nu$ in Eq. (\ref{eq15a}) does not contain $\Delta$ since its  dependence on $\Delta$ is nonlinear. { In order to understand the relative importance of  spin-orbit coupling ($\alpha_s$) and helicoidal gauge potential ($\beta$) on JT oscillation, we define a parameter $s=\alpha_s-\beta$ and plot $\nu$  versus $s$ in Fig. 3 for non-uniform SOC by taking unequal values of $\alpha_s$ and  $\beta$. We see that  $\nu$ decreases with the increase of $s$}.   This indicates   that the spin-orbit coupling and helicoidal gauge potential oppose each other in changing population imbalance (left panel).  The frequency of JT oscillation  with $s$ in presence and absence of NOL are different (right panel of Fig. 3)  and their difference ($\delta \nu$)  decreases as $s$ increases.
\begin{figure}[h!]
\includegraphics[scale=.25]{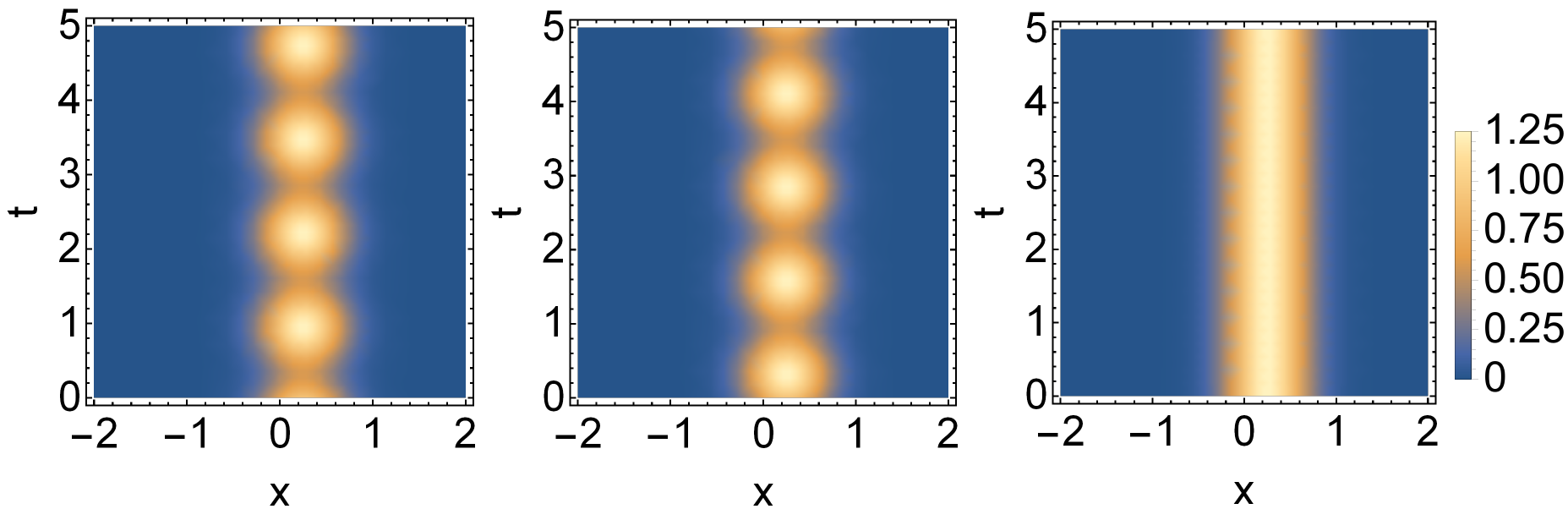}
\caption{Density evolution of the first component $\rho_1=|\psi_1|^2$ (left panel), second component $\rho_2=|\psi_2|^2$ (middle panel) and total $\rho$= $\rho_1+\rho_2$ (right panel) for $\beta=1.6$ and $\varphi(0)=\pi/8$. Other parameters are kept same with those used in Fig. 3.}
\label{fig4}
\end{figure}

Variations of atomic densities with time during JT oscillation in the two states of the system are shown in Fig. 4. It shows that the density of atomic population is changing between the two components (left and middle panel) periodically with time and the change of population imbalance is symmetric. This implies that maximum population in one component corresponds to minimum population of other component such that the total population remains constant with time (right panel).
 
\section{Effects of helicoidal gauge potentials and optical lattices on self-trapping}
In the case of JT oscillation, we have seen that population imbalance ($Z(t)$) oscillates with time about $Z=0$. Understandably, the oscillation of $Z$ about zero imbalance depends on the values of the different parameters of the system. We notice that there exists critical values of the parameters for observation of JT oscillation beyond which the oscillation does not occur about $Z=0$ and thus an asymmetry in the maximum population imbalance arises between the two states. Due to the asymmetry, the average value of population imbalance over one period becomes non-zero. {This is termed as  quantum mechanical self-trapping~(QMST) \cite{r48}}. In order to derive a condition for the occurrence of self-trapping, we recast Eqs. (11) and (12) in terms of Hamiltonian equations of motion, $\dot{Z}=-\partial H/\partial\varphi$ and $\dot{\varphi}=\partial H/\partial Z$ with the Hamiltonian given by
\begin{eqnarray}
H = \frac{(\Lambda+B)Z^2}{2}-A\sqrt{1-Z^2}\cos\varphi-\Delta Z.
\label{eq19}
\end{eqnarray}
For fixed values of initial population imbalance and phase, the macroscopic self-trapping takes place if $H(Z(0),\varphi(0))$ exceeds the larger energy of the system \cite{r48}. Therefore, the critical condition for the observation of self-trapping can be obtained from Eq. (15) as
\begin{eqnarray}
\chi_c > \frac{2}{Z(0)^2}\big[1+\sqrt{1-Z(0)^2}\cos\varphi(0)+\frac{Z(0)\Delta}{A}\big].
\label{eq20}
\end{eqnarray}
Here, $\chi_c=(\Lambda + B)/A$. We see that, for a given value of $\Delta$, $Z(0)$ and $\varphi(0)$, the lattice and SOC parameters can be chosen properly for the observation of self-trapping. The inequality in Eq. (16) is termed as macroscopic quantum mechanical self-trapping~(MQST) condition.

Variation of population imbalance with time for chosen values of different parameters satisfying the MQST condition is shown in the left panel of Fig. 5.  It shows that the population imbalance oscillates about a non-zero value and the amplitude of oscillation increases with the decrease of helicoidal gauge potential ($\beta$) (left panel). This implies that the self-trapping increases with the increase of halicoidal gauge potential. In order to quantify the amount of self-trapping, we calculate average value ($\langle Z(t)\rangle$) of population imbalance over a period for different values of $\beta$ and see that $\langle Z(t)\rangle$ increases nonlinearly as $\beta$ increases(right panel). 

\begin{figure}[h!]
\includegraphics[scale=.34]{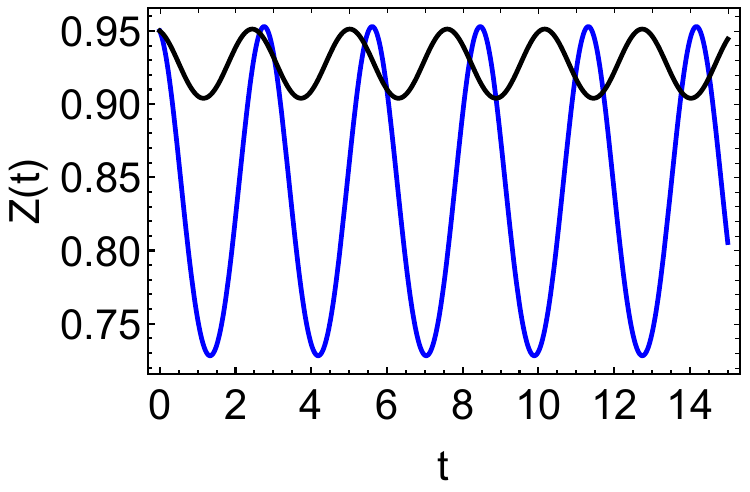}
\includegraphics[scale=.29]{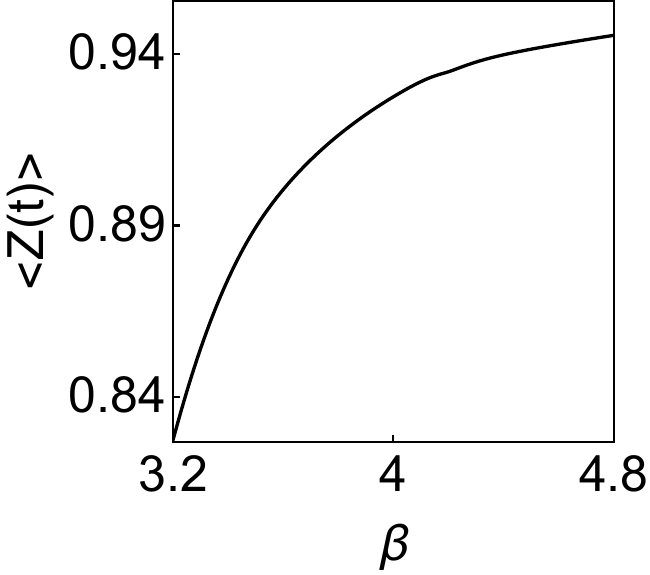}
\caption{Left panel: Variation of population imbalance with $t$ for $\beta=3.2$ (solid blue) and $\beta=4$ (solid black). Right panel: It gives $\langle Z(t)\rangle$ versus $\beta$. In both the panels, we take  $Z(0)=0.95$, $\Lambda=1.817$, $\alpha_s=4$, $x_0=0.25$, $G_-=2.05$, $k_l=5$, $k_n=1.8$ and $\Delta=0$.}
\label{fig5}
\end{figure}

We examine the effects of interaction parameter $\Lambda$ through the variation of population imbalance with time by taking  $\Lambda= 0.665$ (solid blue) and $\Lambda=1.26$ (solid black) in Fig. 6.  The value of $\Lambda$ are so chosen that the binary condensates remain miscible. We see that the amplitude of oscillation decreases with the increase of $\Lambda$. This indicates that the self-trapping is influenced by the relative value of inter- and intra-atomic interactions. We calculate $\langle Z(t)\rangle$ for different values of $\Lambda$ and see that  the self-trapping  grows linearly up to $\Lambda\approx 1.2$.

\begin{figure}[h!]
\includegraphics[scale=.33]{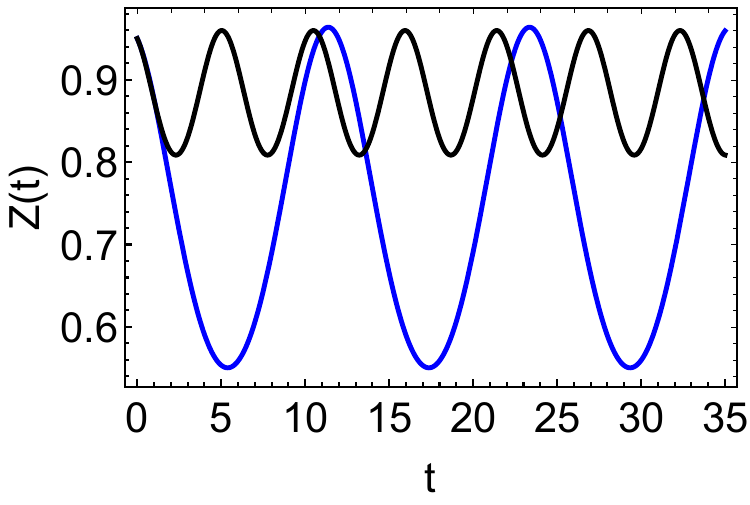}
\includegraphics[scale=.28]{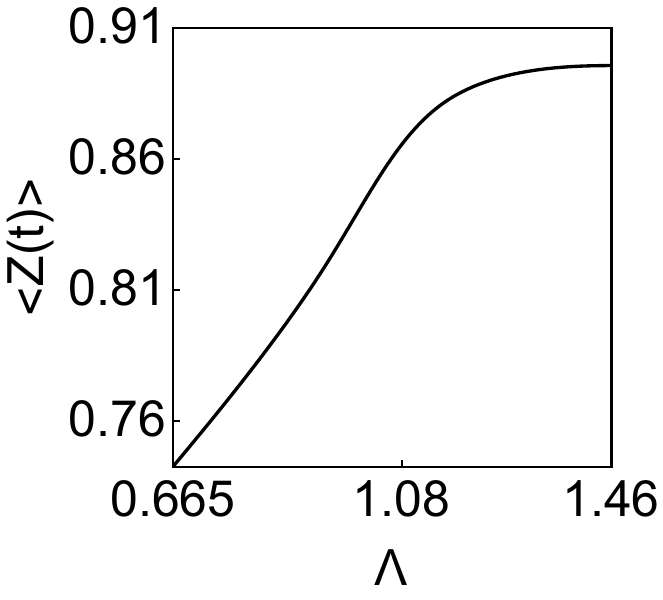}
\caption{Left panel: Variation of population imbalance with $t$ for $\Lambda=.665$ (solid blue) and $\Lambda=1.26$ (solid black). Right panel: It gives $\langle Z(t)\rangle$ versus $\Lambda$. In both the panels, we take $Z(0)=0.95$, $\beta=1.6$, $\alpha_s=4$, $x_0=0.25$, $G_-=2.05$, $k_l=5$, $k_n=1.8$ and $\Delta=0$.}
\label{fig6}
\end{figure}

In order to find the effects of nonlinear optical lattices on the self-trapping, we plot $Z(t)$ for two different lattice strengths, namely, $G_-=2.05$ (solid blue) and $G_-=3.05$ (solid black) in Fig. 7 (left panel). We know that the nonlinear optical lattice influences the phase velocity of $Z(t)$ oscillation  and thus plays an important role in changing the frequency of the oscillation \cite{r20}. {The variation of population imbalance for different values of NOLs' strengths shown in the left panel of Fig. 7 clearly indicates that $Z(t)$ oscillates with time and the amplitude of oscillation increases as the difference of lattice strengths ($G_-$) decreases. The population imbalance in this case also oscillates about $Z(t)\neq 0$ indicating self-trapping.} We calculate $\langle Z(t)\rangle$ for different values of $G_-$. The result displayed in Fig. 7 (right panel) shows that the self-trapping first increases and then attains a saturated value if $G_-$ further increases. 

{In the above discussion we take zero Zeeman splitting. With a view to check the effects of Zeeman splitting ($\Delta$), we plot in Fig. 8, the variation of population imbalance with time for two different values of $\Delta$. It is seen that amplitude of oscillation increases while the frequency deceases for non-zero Zeeman spliting (left panel).  The values of $\langle Z(t)\rangle$ decreases as $\Delta$ increases (right panel). This implies that the self-trapping decreases if we take as $\Delta\neq 0$ for particular choices of lattice and SOC parameters.}

\begin{figure}[h!]
\includegraphics[scale=.34]{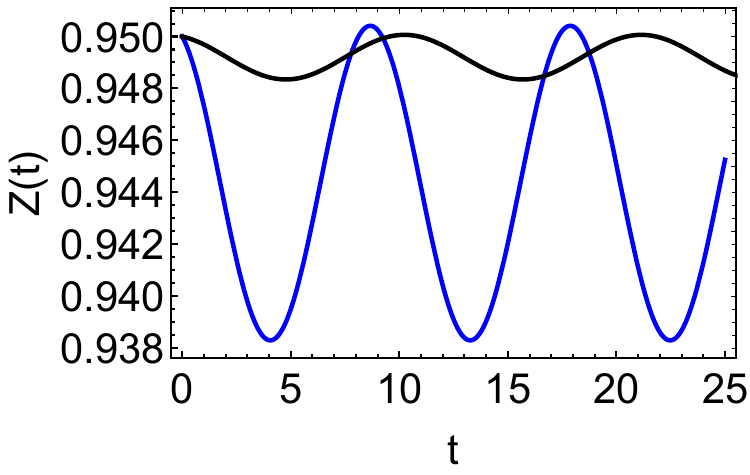}
\includegraphics[scale=.30]{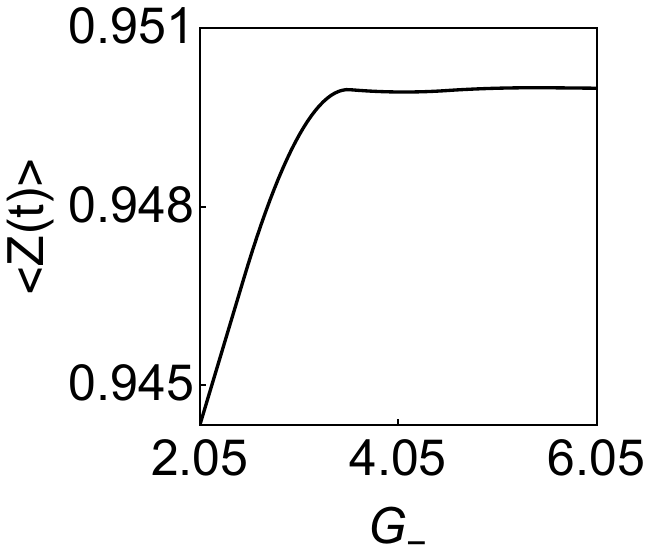}
\caption{Left panel: Variation of population imbalance with $t$ for $G_-=2.05$ (solid blue)and $G_-=3.05$ (solid black). Right panel: It gives $\langle Z(t)\rangle$ versus $G_-$. In both the panels we take  $Z(0)=0.95$, $\beta=2$, $\alpha_s=2$, $x_0=0.25$, $k_l=5$, $k_n=1.8$, $\Lambda=0.681$ and $\Delta=0$.}
\label{fig7}
\end{figure}

\begin{figure}[h!]
\includegraphics[scale=.35]{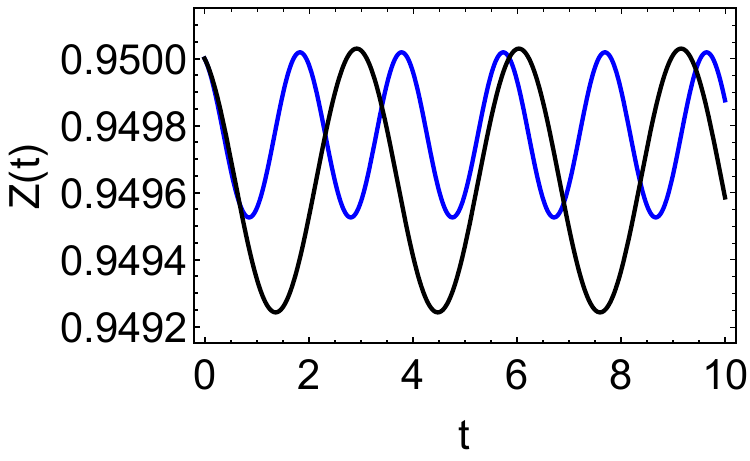}
\includegraphics[scale=.30]{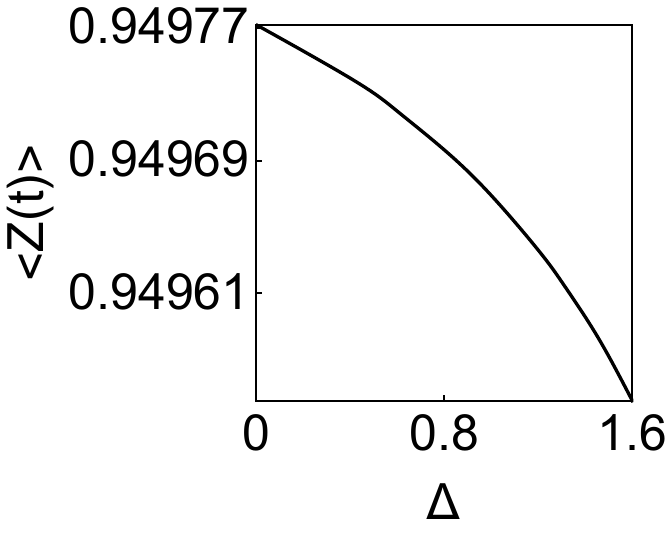}
\caption{Left panel: Variation of population imbalance with $t$ for $\Delta=0$ (solid blue) and $\Delta=1.2$ (solid black). Right panel: It gives $\langle Z(t)\rangle$ versus $\Delta$. In both the panels we take $Z(0)=0.95$, $\beta=5$, $\alpha_s=0.2$, $x_0=0.25$, $k_l=5$, $k_n=0.8$, $\Lambda=1.81$ and $G_-=2.05$.}
\label{fig8}
\end{figure}

\begin{figure}[h!]
\includegraphics[scale=.23]{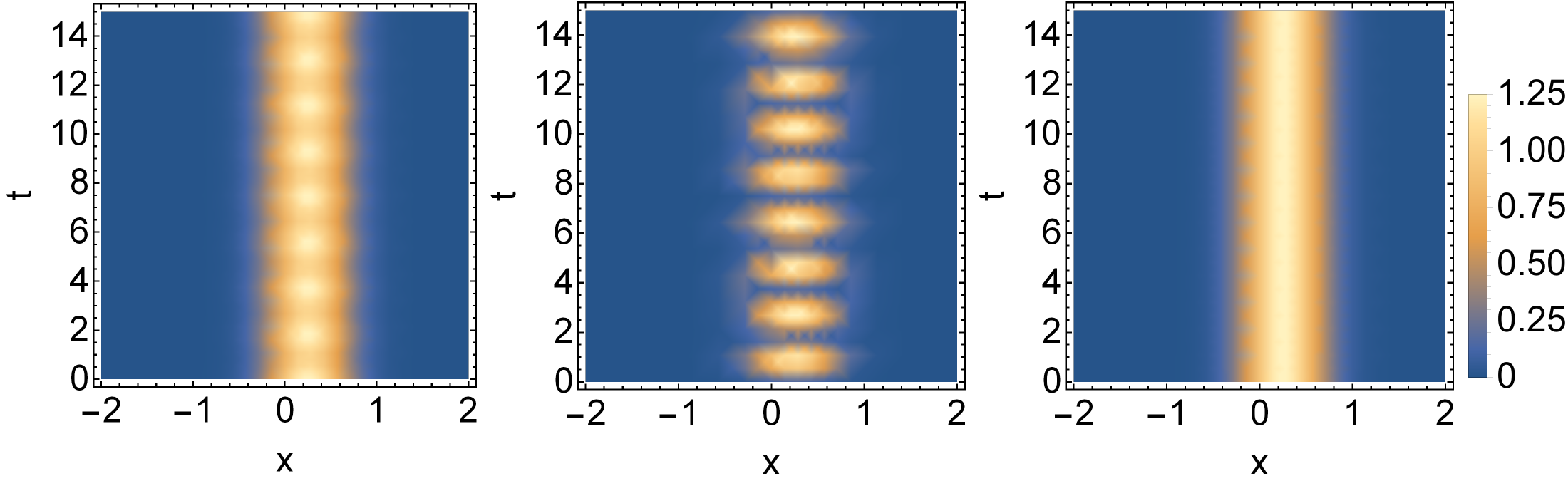}
\caption{Density evolution of the first component $\rho_1=\lvert\psi_1\lvert^2$ (left panel), second component $\rho_2=\lvert\psi_2\lvert^2$ (middle panel) and total $\rho$= $\rho_1+\rho_2$ (right panel) for $\alpha_s=4$, $x_0=0.25$, $\beta=3.2$ and $\Delta=0$.}
\label{fig9}
\end{figure}

{The variation of atomic density ($\rho_j$=$\lvert\psi_j\lvert^2$) with time during quantum mechanical self-trapping is displayed in Fig. 9. It shows that the maximum and minimum densities of the second component are clearly distinguishable while they are not clearly distinguishable in the first component.  This implies that most of the atoms are trapped in the first component and maximum or minimum of atomic density are not clearly visible. However, a few of atoms are involved in the exchange processes and thus causing density variation in the second component.}

\section{conclusions}
We consider a helicoidal spin-orbit coupled Bose-Einstein condensate embedded in optical lattices and studying the time evolution of population imbalance between two pseudo-spin 1/2 states.  More specifically, we consider {two important ingredients} for the system, namely, the effects of (i) helicoidal gauge potential and (ii) nonlinear optical lattices on population imbalance in a spin-orbit coupled Bose-Einstein condensate. It is seen that the population imbalance oscillates in time about a point with zero population imbalance. The amplitude of oscillation depends on the initial phase difference between the two states. In close analogy with the phase dependent current between two superconductor separated by a thin layer we refer the oscillation of atomic imbalance between spin-orbit coupled states as Josephson-type~(JT) oscillation. We have seen that the frequency of oscillation increases as helicoidal gauge potential becomes weaker. In presence of nonlinear optical lattices, the frequency of Josephson-type oscillation for a fixed helicoidal gauge potential further increases.

We have examined that the oscillation of population imbalance does not occur about zero imbalance if the parameters of the system do not satisfy the so-called macroscopic quantum mechanical self-trapping condition. In order to quantify  self-trapping, we calculating average value of population imbalance for different values of (i) helicoidal gauge potential, (ii) nonlinear lattice parameters and (iii) inter and intra-atomic interaction strengths. We find that, for a particular value of helicoidal gauge potential, self-trapping grows appreciably with the change of atomic interaction and lattice parameters. However, we observe saturation in the self-trapping for higher values of atomic interactions. {Thus we conclude by noting that  both the BECs with  uniform and non-uniform SOCs can exhibit oscillation of atomic population imbalance and self-trapping. However,   frequency of oscillations decreases as the system approaches towards the non-uniform SOC BECs. We also observe that self-trapping can be increased  as non-uniformity in SOC increases.}  

\section*{Acknowledgements}
S. Sultana would like to thank  `West Bengal Higher Education Department' for providing Swami Vivekananda Merit Cum Means Scholarship with F. No. WBP201637916084.

\end{document}